# Enhanced Asymmetrically-Clipped Optical OFDM

**Arthur James Lowery**

*Electro-Photonics Laboratory, Department of Electrical and Computer Systems Engineering, Monash University, VIC 3800, Australia*
*arthur.lowery@monash.edu*

**Abstract**: Asymmetrically clipped optical orthogonal frequency-division multiplexing (ACO-OFDM) is a technique that sacrifices spectral efficiency in order to transmit an orthogonally frequency-division multiplexed signal over a unipolar channel, such as a directly modulated direct-detection fiber or free-space channel. Several methods have been proposed to regain this spectral efficiency, including: asymmetrically clipped DC-biased optical OFDM (ADO-OFDM), enhanced U-OFDM (EU-OFDM), and spectral and energy efficient OFDM (SEE-OFDM). This paper presents a new method that offers the highest receiver sensitivity for a given optical power at spectral efficiencies above 3 bit/s/Hz, having a 7-dB sensitivity advantage over DC-biased OFDM for 1024-QAM at 87.5% of its spectral efficiency, at the same bit rate and optical power.

**Author's Note 7th Nov. 2015**

A recent post of Arxiv on a modified SEE-OFDM (*Emily Lam, Sarah Kate Wilson, Hany Elgala, Thomas D. C. Little*, "Spectrally and Energy Efficient OFDM (SEE-OFDM) for Intensity Modulated Optical Wireless Systems," 28 Oct. 2015, arXiv.org > cs > arXiv:1510.08172) has drawn my attention to *Qi Wang, Chen Qian, Xuhan Guo, Zhaocheng Wang, David G. Cunningham, and Ian H. White* "Layered ACO-OFDM for intensity-modulated direct-detection optical wireless transmission" (Optics Express, 4 May 2015, Vol. 23, No. 9, p. 12382, https://www.osapublishing.org/oe/abstract.cfm?uri=oe-23-9-12382), which uses the same harmonic sequences as my proposal here, and also a method of reconstructing the clipping noise from the estimated constellation values. My work herein presents more comparisons on the relative performance of EACO-OFDM versus other techniques particularly DCO-OFDM, and also up to the high levels of spectral efficiency, where I have identified the greatest performance gains. I have also verified that the power penalty over a single ACO channel can be accurately calculated by considering the DC levels of individual chords/layers of the ACO sub-channels. My comments on SEE-OFDM in the Introduction refer to the original paper [17], and not the latest Arxiv version (28th Oct. 2015).

**Author's Note 9th Mar. 2016**

Arthur James Lowery, "Comparisons of spectrally-enhanced asymmetrically-clipped optical OFDM systems," Opt. Express 24, 3950-3966 (2016) https://www.osapublishing.org/oe/abstract.cfm?uri=oe-24-4-3950

Note also that there is a more recent paper by Harald Haas's group that also proposed a layered ACO-OFDM scheme, coincidently called Enhanced ACO OFDM:

Mohamed Sufyan Islim, Dobroslav Tsonev and Harald Haas, "On the Superposition Modulation for OFDM-based Optical Wireless Communication" GlobalSIP, Orlando, Florida, Dec. 2015. Session: WbOW-L: Modulation and Coding, Paper 1.

*Thus this is a very active field with important outcomes for improving the performance of intensity-modulated optical communications systems.* Please search for more recent papers, regularly.

## 1. Introduction

Asymmetric clipping of optical orthogonal frequency-division multiplexed (O-OFDM) channels [1, 2] can reduce the optical power requirement for a given data rate because no optical power is wasted in DC-biasing a bipolar OFDM waveform to become unipolar – negative values are simply set to zero. This zero-added-bias O-OFDM offers a greater electrical signal to noise ratio (SNR) for a given optical power [3] than simply adding a large DC-bias as in Direct-Detection O-OFDM (DDO-OFDM) to avoid clipping of the negative-going peaks. A high SNR can support high-order quadrature amplitude modulation (QAM), increasing the attainable data rates in fiber and wireless light communications systems that have a limited modulation bandwidth of their electro-optic components. Alternatively coherent optical channels can provide high SNRs [4], including channels with a reference carrier [5]; however, intensity modulation is simpler and less-bulky, and importantly is lower cost for applications such as for visible lightwave communications [6] and short-reach fiber links.

Common forms of clipped OFDM signal include: asymmetrically clipped optical OFDM (ACO-OFDM), which uses spectral allocation of subcarriers to reduce or eliminate the effects of clipping (either a gap between the subcarrier band and DC [1], or by using only odd subcarriers [2]), and Flip OFDM [7, 8] (also known as unipolar OFDM, or U-OFDM [9]), which transmits the positive, then negated negative, parts of the signal in two consecutive positive-valued blocks. ACO-OFDM and Flip/U-OFDM offer the same spectral efficiencies, due to either the halving of the number of available subcarriers (ACO-OFDM) or the doubling of the required time to transmit an OFDM symbol (Flip/U-OFDM). However, at the receiver, Flip/U-OFDM de-maps multiple frequencies in the transmitted channel into each optical subcarrier, so it is difficult to equalize the channel's imperfections in the frequency domain. Due to half the transmitted samples being zero-values, techniques are available to reduce noise in Flip/U-OFDM, using information in the clipped portion of the waveform [7]. Similar techniques exist for ACO-OFDM [10-13]. Flip/U-OFDM systems offer advantages in the size of the Fourier transform, which can be halved compared with ACO OFDM.

To improve spectral efficiency, it is possible to superimpose a DCO-OFDM channel upon an ACO-OFDM channel, using the odd-frequency subcarriers for ACO-OFDM and the even-frequency subcarriers for DCO-OFDM – called asymmetrically clipped DC-biased optical OFDM (ADO-OFDM) [14]. The DCO-OFDM channel can be recovered by first cancelling the distortion (and signal) of the ACO-OFDM channel, to reveal the DCO-OFDM signal. The cancellation waveform is derived by: (1) using a fast Fourier transform on each OFDM symbol to separate all of the subcarriers, (2) selecting the odd subcarriers, (3) using an inverse fast Fourier transform then clipping to recreate close to transmitted ACO-OFDM waveform from the selected subcarriers. This scheme gives optimal performance when a lower constellation size is used on the DCO-OFDM subcarriers than the ACO-OFDM subcarriers, but has the disadvantage that the cancellation process degrades the DCO-OFDM signal by 3-dB [14]. A similar scheme uses ACO-OFDM on the odd subcarriers, and a Flip/U-OFDM type signal on the even subcarriers at only one-half of the baud rate of the ACO-OFDM signal [15].

Recently, Enhanced-UOFDM [16] (EU-OFDM) has been proposed to increase the spectral efficiency of Flip/U-OFDM, again using a cancellation technique. EU-OFDM uses the superposition of time-domain signals at various 'depths', where each symbol in deeper layers are transmitted more than once to enable their cancellation at the receiver; for example, depth $d$ is transmitted $d$ times in succession, lowering its spectral efficiency by $1/d$, but improving the overall spectral efficiency of the system. Slicing, to recover the transmitted bit values, is used to remove the noise from shallow depths during the cancellation process; however, EU-OFDM places the subcarriers of all depths at the same frequencies, which unfortunately means that any errors in slicing the constellations of the lower depths results in strong error vectors at the deeper depths, causing a cascade of bit errors. A further issue with Flip/U-OFDM is that the positive- and negative-signal blocks must not interfere with one another in a bandwidth-limited channel. This either requires time-domain equalization before the processing of the blocks, or additional cyclic prefixes between the blocks [16]. A similar scheme—spectral and energy efficient (SEE) OFDM—uses a combination of one ACO-OFDM path for the odd subcarriers, and Flip-OFDM processing of ACO-OFDM generated with half/quarter-size transforms on further paths to in-fill the even subcarriers [17]; this has the advantage that only one FT is required at the receiver, and also that all paths have the same overall symbol duration. However, slicing is not used, so noise can transfer from one set of subcarriers to other sets during cancellation. Also, four cyclic prefixes are required within this symbol, increasing the cyclic-prefix overhead, unless longer symbols (and so longer transforms, with more latency) are used.

This paper presents a frequency-domain method of generating and receiving ACO-OFDM channels so that the overall spectral efficiency is increased while still maintaining the unipolar nature of the signals and the clipping-at-zero technique to remove the need for biasing. Similar to SEE-OFDM, this technique relies on successive filling of the even subcarrier frequencies of ACO-OFDM; however, it uses additional ACO-OFDM channels without any Flip-OFDM processing. Thus, all of the subcarriers are periodic within a common symbol duration, so can share one common cyclic prefix per symbol. This lowers the potential CP overhead compared with SEE-OFDM and EU-OFDM. Using ACO-OFDM also enables frequency-domain processing of imperfect channels and bit loading. As with these other methods, at the receiver, the channels are successively decoded, starting with the original odd-subcarrier channel. In contrast to ADO-OFDM and SEE-OFDM, slicing (thresholding) is used for each decoding stage, and for the generation of the cancelling waveforms. Thus, for reasonable SNRs, the penalty due to imperfect cancellation of a channel is very small, as there is little noise on the cancellation waveform, and whatever sporadic noise there is (due to an incorrect decision, for example), is spread over several subcarriers, unlike with EU-OFDM.

Unlike, EU-OFDM, EACO-OFDM assigns each subcarrier to a unique frequency, so that slicing errors during the cancellation process do not lead to cascades of bit errors.

A simple analysis of the fundamental optical-power penalty of superimposing separately clipped ACO-OFDM channels is provided, and agrees with the simulation results. This indicates that EACO-OFDM offers most improvement over DCO-OFDM at high signal to noise ratios, in a similar way to the characteristics of the other schemes. However, because slicing is used in the cancellation process, noise is not strongly propagated between depths, this improvement can be substantial. Because each subcarrier can be frequency-domain equalized to further improve signal quality, EACO can support higher-order QAM signaling, enabling high data rates on bandwidth-limited optical-OFDM channels. Numerical simulations show that for 1024-QAM this *enhanced* ACO-OFDM (EACO-OFDM) offers 7-dB better performance than DCO-OFDM at the same optical power, data rate and receiver noise, with a loss of only 12.5% spectral efficiency.

## 2. Principle of EACO-OFDM

### 2.1 Odd-subcarrier ACO-OFDM

The technique is based on odd-subcarrier ACO-OFDM, as illustrated in Fig. 1 [1]. Odd subcarriers are generated by using only half the available frequencies, by providing, e.g. QAM data to only the odd inputs of the inverse Fourier transform (IFT) within the required transmission band. Also, only the real *outputs* of the IFT are used, which means that computational efficiency can be gained by using an IFT designed to produce only real-valued outputs [18]. This is an alternative to forcing the IFT to produce no power at its imaginary valued outputs by using positive and negative frequency input pairs with Hermitian symmetry.

The output of the IFT is converted to a serial stream of OFDM symbol blocks using a parallel-to-serial (PS) converter. The serialized samples are converted to an analog waveform by a digital-to-analog converter (DAC). The clipping is illustrated after the DAC, but could be before it. A gain of 2× can be used to restore the subcarriers to the same amplitude as a DCO-OFDM system [14]; however, the energy per bit calculation should include the associated power in the distortion products and DC power that are created during clipping. The even-frequency subcarriers accommodate the intermodulation distortion that is created when the bipolar OFDM signal is clipped at its mean level, i.e. when negative-valued signal samples are set to zero. As illustrated in Fig. 1, intermodulation distortion products are also generated at higher frequencies than the signal chord. The clipping also produces a DC component to the signal spectrum, which is translated to a mean optical power. This can be thought of as 'automatic self-biasing', as it is always at the optimum level, without the need for a control circuit. Importantly, as we shall see in the penalty calculations, the level of the mean DC component increases with the sum of the square-roots of the numbers of used subcarriers within each chord.

The optical channel model is usually represented at baseband in intensity-modulated systems with broad optical spectra compared with the modulation bandwidth. Generally, the signal voltage at the transmitter is mapped onto optical intensity, for example, by modulating the current driving a LED in proportion to the signal voltage. At the detector, the optical intensity is converted into a photocurrent, which is then converted to a voltage using a trans-impedance amplifier (TIA). The voltage-out to voltage-in of this channel and detector is approximately linear for low levels of modulation, though the effects of device nonlinearity should be considered in order to realize practical systems. The frequency response of the photodiode and associated electronics can be represented by a low-pass filter.

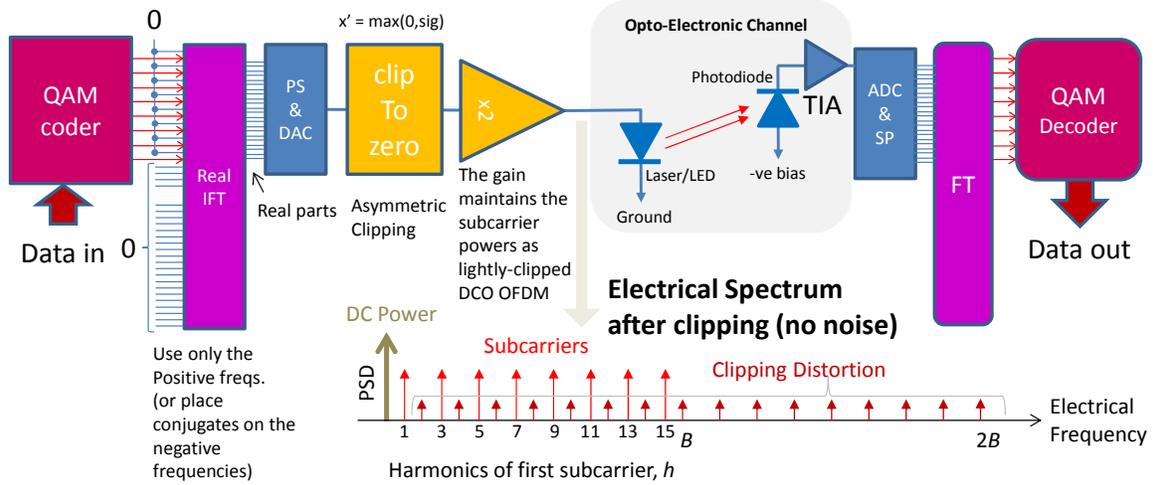

Fig. 1. Odd-subcarrier ACO-OFDM system block diagram. Top: system block diagram. Bottom: illustrative spectrum after the clipping process.

At the receiver, the voltage output of the TIA is digitized with an analog to digital converter (ADC). Noise in the photodiode, and from the TIA and digitizer, adds to the signal samples. These noise sources will be considered as additive white Gaussian noise (AWGN) added to the photocurrent. A serial-to-parallel (SP) converter feeds blocks of samples, corresponding to an OFDM symbols, to a Fourier transform (FT), which acts as matched filters for the subcarriers. The odd subcarriers are free from clipping distortion and so are decoded in the QAM decoder. The even-frequency FT outputs are usually discarded, even though they could carry useful information, because they are created from the wanted subcarriers using a deterministic process. The signal quality benefit of such processing is not discussed in this paper, because the even-frequency subcarriers are used to carry additional signals.

*2.2 Enhanced ACO-OFDM transmitter*

A key enabler for EACO-OFDM is that the clipping distortion in an ACO-OFDM signal is deterministic and only falls on even subcarriers. It can therefore be recreated from the signal in the odd subcarriers, then mostly cancelled at the receiver. This cancellation allows additional ACO-OFDM subcarriers, placed at the even subcarrier frequencies, to be recovered, by first cancelling the clipping distortion. This observation was also used in ADO-OFDM [14] and SEE-OFDM [17].

A musical analogy will be used to describe the subcarrier allocations in the frequency domain. This is appropriate, because each part of the transmitter generates a harmonic sequence of subcarriers, similar to a musical *chord*; however, each part's chord is doubled in frequency, which is a musical *octave* from the previous chord. As in music, notes of one chord (*i.e.* the subcarriers from one part of the transmitter) interleave with notes from other chords, rather than being contained within discrete bands of frequencies.

The subcarrier allocations (*notes*) are arranged so that the receiver is able to cancel the clipping distortion from lower chords in order to recover the higher chords with perfect fidelity. At the transmitter, each chord is asymmetrically clipped *before* being combined with the other chords. As shown in Fig. 1 (left-upper), the lowest chord, C0, uses only odd harmonics of its lowest subcarrier (musically, e.g.: harmonic/note: 1=**C**, 3=**G'**, 5=**E''**…). A chord, C1, at double the frequency (musically, an octave above the first) will fill some of the even frequencies of C0 (2=**C'**, 6=**G''**, 10=**E'''**…), and a chord, C3, at quadruple the frequency will fill yet more even subcarriers (4=**C''**, 12=**G'''**, 20=**E''''**…). The spectral utilization is therefore ½ + ¼ + ⅛ = ⅞, compared with ½ for ACO-OFDM. Although not explicitly stated, Elgala and Little's system [17] uses the same sequence for its higher 'paths', because the ACO-OFDM generates odd frequencies, and using ½ and ¼-sized transforms multiples these frequencies by two and four.

The process in Fig. 2 will increase the spectral efficiency to 87.5%. Generally, if the chords are labelled $c = 0, 1, 2\ldots$, and the odd harmonics within a chord $h = 1, 3, 5 \ldots$, the relative frequency of a subcarrier will be $2^c \times h$. For an OFDM symbol of duration $T_{OFDM}$ (before cyclic prefix addition) the frequencies of the subcarriers will be $(2^c \times h)/T_{OFDM}$.

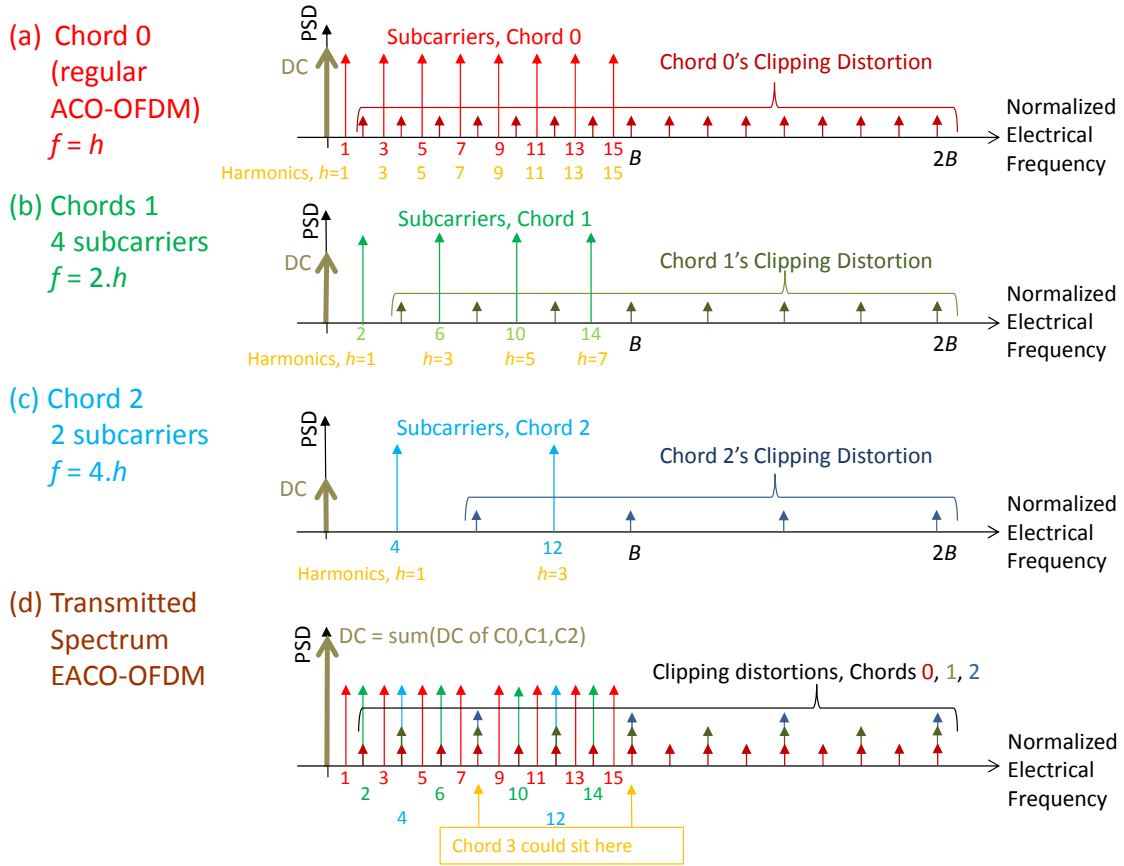

Fig. 2. Process for generating an Enhanced-ACO-OFDM signal at the transmitter, by adding together the clipped waveforms of several chords, illustrated in the spectral domain. Note that there are only 16 possible subcarriers in this example (14 are used, in 3 chords), for clarity: the simulations will use 56 out of 64 possible subcarriers. The frequencies are normalized so that the subcarrier separation in Chord 0 is unity, implying a unity-duration OFDM symbol.

### 2.3 Enhanced ACO-OFDM receiver: cancellation process

At the receiver, the higher chords must be substantially cancelled to reveal the lower chords. The cancellation process for C0 is shown in Fig. 3, along with illustrative spectra. This process is used for each successive OFDM symbol on a symbol-by-symbol basis. Firstly, the frequency domain information is extracted from the received signal using a serial-to-parallel converter and a real-input Fourier transform. Secondly, the information-bearing subcarriers corresponding to C0 (and any other signal that falls at these frequencies) are sliced (thresholded) to recover an estimate of the QAM modulation. Then, the time-waveform is recreated using an inverse Fourier transform. This waveform is zero-clipped then subtracted from the received signal. The spectrum at the bottom of Fig. 3 illustrates that the signal and distortion products from C0 have been removed – C1 is now free of pollution from the signal of C0 (including its clipping products). The QAM symbols could also be extracted from the C0 subcarriers after the slicer. To recover C2, a second iteration of cancellation is required, using a similar arrangement; this time to recreate the signal and distortion products of C1. Because ACO-OFDM is used on all chords, and not just C0, the functional descriptions of the cancellation processes in each iteration are almost identical.

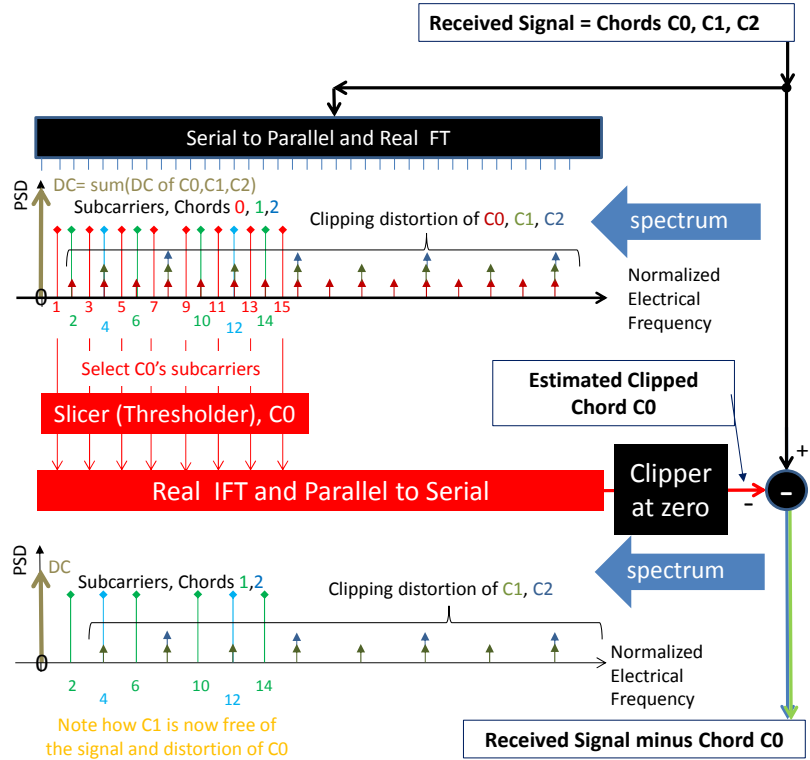

Fig. 3. Process for cancelling the signal and distortion products from Chord 0, enabling Chord 1 to be processed. The QAM symbols in C1 can also be extracted after the slicer.

For the receiver to work effectively, the chords are generated and clipped independently at the transmitter: the technique would not work if the entire bandwidth was occupied (as in DCO-OFDM), and then clipped as one before transmission. This is because, using C0 as an example, there must be a definite correspondence between the clipping distortion and the signal in C0, so that both can be cancelled at the receiver. If chords C0 and C1 were clipped after superposition, then the clipping distortion of C0 could not be cancelled to reveal C1, because the signal of C1 would have influenced which samples would have been clipped.

Importantly, noise added to the received signal will mean that the slicing (thresholding) operation will introduce errors into the estimate of a given chord's signal and distortion. As we shall show, this is only a significant issue at low signal to noise ratios; importantly, it means that there only is around 0.8-dB when reconstructing the highest chords, for a 1024-QAM system designed to give a BER of $10^{-3}$.

## 3. Simulations

A 3-chord EACO-OFDM system was compared to a single-band DCO-OFDM system using simulations from VPItransmissionMaker™ version 9.1. The Fourier transform lengths were $N_{FFT}$=1024 points, with all chords occupying up to the first 64 subcarriers, excluding DC. This gives 16-times oversampling in the simulations. The OFDM symbol rate was 10 Gsymbol/s, thus the simulation sampling rate was $R_{sim}$ = 160 Gsamples/s, giving an electrical simulation bandwidth was 80 GHz, to give room for distortion products to fall into. 4-QAM (QPSK) was first used for modulation, giving a data rate of 2×10G×(32+16+8)/64 = 17.5 Gbit/s. Of course, the results are easily scaled to other data rates and also increase with higher-order QAM. A total of 256 OFDM symbols were simulated for each data point. No clipping was applied to any positive excursions of either the ACO-OFDM or the DCO-OFDM system, to allow comparisons with previous publications on enhancing spectral efficiency.

The transmitted signals (DCO- and EACO-OFDM) were fed through optical power-control circuits. The assumption is that the optical power is proportional to the signal voltage. The power control works by calculating the mean value of the voltage signal over 32 OFDM symbols, then dividing all signal samples by this value. Bandwidth-limiting and nonlinearities of the optical channel, including the source and the photodiode, were not modeled, as with previous papers on spectral efficiency improvement.

The ACO-OFDM receiver's noise was added to the photocurrent at the receiver. The noise level at the receiver was set so that a standard odd-subcarrier ACO-OFDM system gives an error vector magnitude (EVM) of 0 dB for an indicated SNR of 0 dB for 4-QAM. After the FTs, the EVM was estimated from the constellations and a knowledge of the transmitted data. Note that the EVM calculation uses the average power per symbol, calculated using all constellation points. For comparison purposes, the DCO-OFDM receiver used exactly the same noise variance as the EACO-OFDN, and the same optical input power.

Note that the EVMs are estimated over all subcarriers in a given chord, and so ignore the fact that some subcarrier frequencies have more distortion products falling upon them than others, which would not be cancelled if there was a slicing error. However, this metric is sufficient to illustrate the advantages of the enhanced ACO-OFDM over other approaches, even though the BER could, in practice, be improved over the value predicted by the $Q$ estimate using techniques such as, for example, pairwise coding [19], constellation optimization and receiver clipping [20], or Walsh-Hadamard coding [21].

*3.1 EACO-OFDM transmitted spectrum*

Figure 4 shows the transmitted spectrum of the EACO-OFDM system. The three individual chords are shown, plus the combined spectrum. Note that the spectra of each subcarrier, when estimated over multiple OFDM symbols, are sinc-shaped. This is more obvious for the chords with sparser subcarriers (i.e. C2, blue). Note that no subcarriers are transmitted at multiples of 1.25 GHz, which reduces spectral efficiency. Six chords would be required to fill all of these gaps, with the computation associated with 5 cancellation stages.

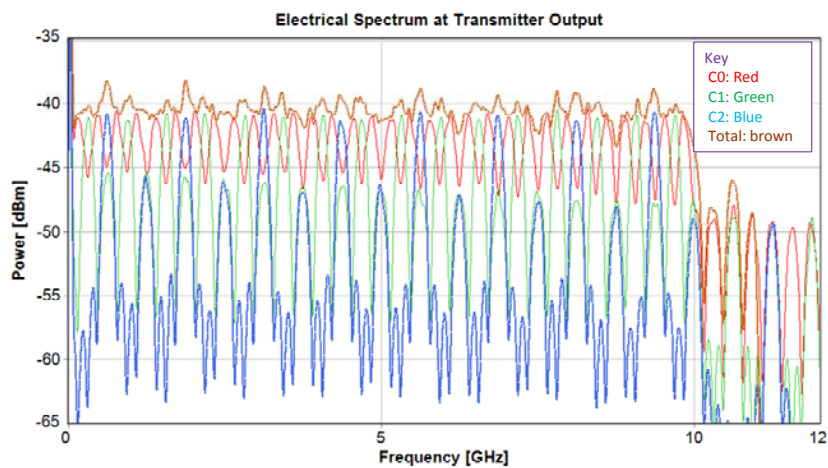

Fig. 4. Electrical spectrum of the transmitted ACO-OFDM signal (brown) with the spectra of Chords 0 (red), 1 (green) and 2 (blue). The spectra are averaged over 8 simulation runs and has a resolution bandwidth of 50 MHz. Note how only 56 out of the 64 possible locations are occupied if only 3 chords are used.

*3.2 Constellation comparisons*

The DC bias of the DCO-OFDM system is critical to its signal quality: if it is set overly high, then too much power is wasted on bias, and less power is used for the signal energy in a subcarrier, lowering its signal quality for a given signal-to-noise ratio. This leads to an EVM penalty of roughly the *Bias* in dB [22]: conversely, if the Bias set to low, the clipping of its negative peaks will lower its performance on lead to a ceiling in signal quality, so that higher-order QAM cannot be supported. For fair comparisons, multiple DC-bias levels will be used in the quantitative comparisons of signal-quality versus SNR later in the paper. In the following comparisons, a Bias of 6.2 dB was used, which is nearly optimum for DCO-OFDM at a SNR of 25 dB. Note that the Bias (dB) is defined in terms of the DC bias, $B_{DC}$, to the r.m.s. a.c. signal voltage (before any clipping), $V_{rms}$, as Bias (dB) = $10\log_{10}(1+(B_{DC}/V_{rms}))$ [22].

Figure 5 shows typical constellations for the EACO-OFDM with the cancellation method of Fig. 3 enabled and disabled. Also shown is the constellation for the DCO-OFDM system. Because no distortion products from higher chords fall on the odd frequencies of C0, C0 is not degraded if cancellation is not used (please compare the red constellations). This could be useful as a fading system can always fall back to using C0 for signaling, whether other chords are being transmitted, in situations where the when the SNR is poor. In contrast, C1 definitely requires the first stage of cancellation in order to be received with a reasonable signal quality; without cancellation C1 suffers at least 14 dB of degradation. With cancellation, its performance is similar to that of C0.

Chord 2 relies even more strongly on cancellation, with over 17-dB improvement in signal quality from cancellation. This is because C2 suffers from distortion products from C0 and, at some frequencies, distortion products from C1. Thus, any estimation errors for C0 and C1 will degrade C2. Overall, the transmission of separately-clipped chords then iterative cancellation at the receiver enables the EACO-OFDM system to provide a signal quality (EVM) at least 3.9-dB better than the DCO-OFDM system, for the same receiver noise, baseband channel bandwidth, number of subcarriers, and optical power.

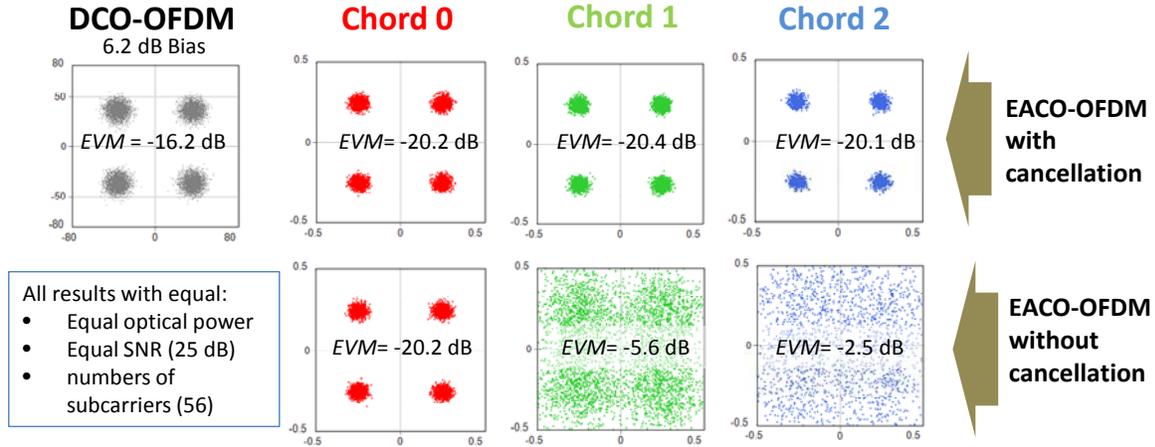

Fig. 5. Simulated constellations for a 3-chord EACO-OFDM system (with and without cancellation) and a DCO-OFDM system biased at 6.2 dB. All have 56-subcarriers and operate at the same received optical power and receiver noise, giving the same SNR. All the chords in the EACO-OFDM system out-perform the DCO-OFDM system, and the necessity of using cancellation to recover chords C0 and C1 is clear.

*3.3 Performance over a range of signal to noise ratios*

Figure 6 plots the *Q*'s of all three ACO-OFDM chords, with five simulation runs of 256-symbols each. Also plotted is the signal quality for DCO-OFDM operating at the same optical power and carrying the same number of subcarriers, for biases of 4 to 10 dB. The DCO-OFDM traces (brown) clearly show that there is an optimum bias if a given signal quality is required. Thus, for fair comparisons, the DCO-OFDM bias level must be set near optimally. Alternatively, a line may be drawn between the 'knees' of each curve, as this used to estimate the signal quality that could be obtained from DCO-OFDM for a given bias. This has a slope of approximately 0.8 dB/dB; that is, an SNR increase of 10 dB gives only an 8-dB increase in signal quality. At high SNRs the DCO-OFDM requires a very strong bias (>9 dB) to give a corresponding benefit in signal quality; however, to support 4-QAM at $10^{-3}$ BER, a bias of only 3 to 4 dB is preferable. Because of the soft-knees of the EVM versus SNR, the exact bias level is not very critical, thus all of the results use near-optimum biasing.

The main observation from Fig. 6 is that all chords of EACO-OFDM have better (lower) EVMs than the optimum DCO-OFDM for SNRs above 13.5 dB. At low SNR's the cancellation technique is imperfect due to errors in the slicing, leading to most degradation in C2. However, at SNRs >15 dB, all three EACO-OFDM chords offer very similar performance. The advantage of EACO-OFDM improves with SNR. This is because at higher SNRs, the EVM of ACO-OFDM improves 1 dB per dB increase in SNR: in contrast, for DCO-OFDM, there is ≈0.8 dB reduction in EVM per dB increase in electrical SNR. Importantly, EACO-OFDM allows higher ultimate signal qualities, if a good SNR is available.

The results in Fig. 6 are for 4-QAM modulation. An open question is whether advantage could be taken of the higher signal qualities available with EACO-OFDM, by using a higher-order modulation scheme. For example, it might be that the higher number of decision thresholds required for, say, 16-QAM, would cause more errors in the estimates, causing imperfect cancellation. Figure 7 plots the EVM for EACO-OFDM and DCO-OFDM using 1024-QAM modulation. The reduction in EVM for EACO-OFDM is far larger at the higher SNRs required to support 1024-QAM, being over 7 dB for 1024-QAM, at EVMs that can support BER <$10^{-3}$. Note that the penalty due to slicing errors *at low SNRs* are far higher than for 4-QAM, giving a 5-dB difference in performance between C0 and C2, compared with 1 dB for 4-QAM. This performance difference has reduced to <0.8 dB at the EVM required for a BER of $10^{-3}$.

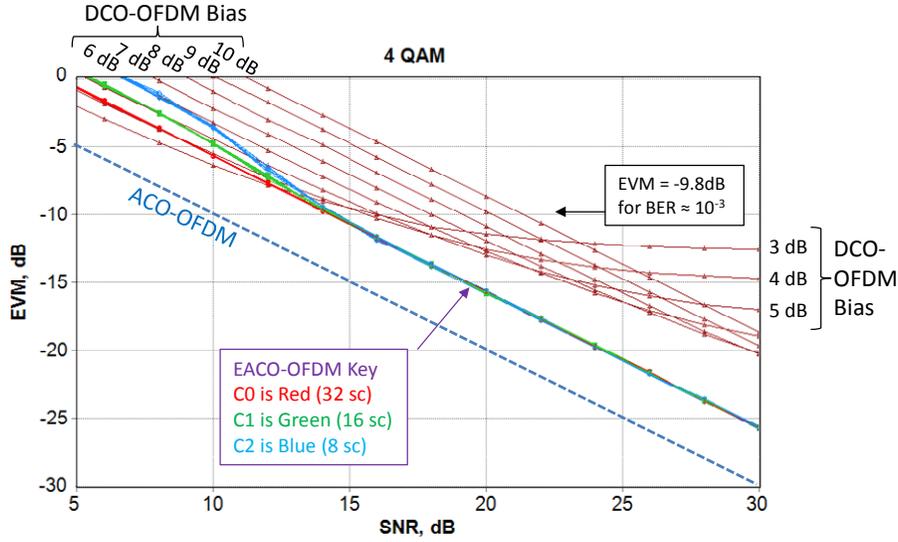

Fig. 6. Signal quality (estimated from constellation spreads) versus signal to noise ratio for 4-QAM EACO-OFDM and DCO-OFDM (8 biases). The dashed line is the signal quality that would be obtained with a (single chord) 56-subcarrier ACO-OFDM system using approximately twice the spectral bandwidth of the other systems. The signal quality of the DCO-OFDM systems is ultimately limited by clipping distortion. Note how the curves for EACO-OFDM converge for high SNRs due to the improbability of errors in the slicing process. Without slicing the separation of the EACO-curves remains approximately constant over all SNRs.

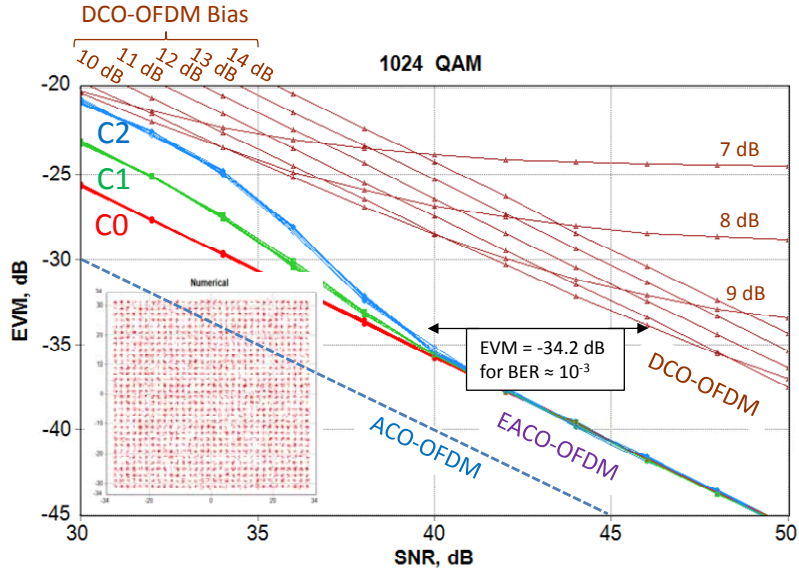

Fig. 7. Signal quality (estimated from constellation spreads) versus signal to noise ratio for 1024-QAM ACO-OFDM and 1024-QAM DCO-OFDM (8 biases). The slicers work very effectively well at BERs better than $10^{-3}$, but not quite as effectively as for 4-QAM.

*3.4 Fundamental power penalty in using more than one chord*

There is a fundamental penalty of using multiple chords. The issue is that for a single chord, the mean optical power required for a given received signal to noise ratio, rises as only the square-root of the number of subcarriers. The power for multiple chords is simply the sum of the powers in each chord. Thus, if we compare a system with 8, 16 and 32 subcarriers in three chords, versus a single chord with 56 subcarriers, the optical power ratio will be ($\sqrt{8}+\sqrt{16}+\sqrt{32})/\sqrt{56} \approx 12.5/7.5 \approx 1\frac{2}{3}$ which is a 2.2-dB increase in mean optical power. In a mean-power-limited intensity modulated system, this implies that the received electrical energy per bit would have to increase by 4.4 dB, so agrees with the simulation results. Therefore, fundamentally, overlaying chords of clipped waveforms to regain spectral efficiency necessarily comes with a penalty. This penalty increases rapidly to over 7.2 dB for 99% spectral efficiency, but asymptotes at 7.7 dB.

The net outcome of the above penalties is that EACO-OFDM works best for higher constellation sizes, where increasing the spectral efficiency by 1.75× would require, say, an increase from say, 64-QAM to 512-QAM, which requires 7 dB of additional SNR. Thus, the 4.4-dB penalty of using 3-chords in EACO-OFDM is obviously preferable.

### 3.5 Comparison of methods of increasing spectral efficiency

Figure 8 plots the simulated SNR cost of increasing the spectral efficiency for equal bit rates and optical powers for a number of different systems. That is, the cost is the amount that the electrical SNR at the receiver would have to increase by to maintain a BER of $10^{-3}$, compared with 4-QAM ACO-OFDM that provides unity spectral efficiency. The results for SEE-OFDM and ADO-OFDM are taken from the literature. Some observations are:

- EACO-OFDM (this paper) offers the lowest cost for SEs higher than 3 bit/s/Hz.

- Pulse Amplitude Modulation (PAM) has the highest cost of all systems, and this cost increases rapidly with spectral efficiency.

- ACO-OFDM [2] has the lowest cost at SEs less than 2 bit/s/Hz; however, its cost increases rapidly with SE, as it has to use an $m^2$-QAM to obtain the same SE as $m$-QAM DCO-OFDM.

- SEE-OFDM [17] offers the lowest cost at an SE of 3 bit/s/Hz.

- ADO-OFDM [14] follows a similar trend to EACO-OFDM, but has a penalty due to noise added to the DCO-OFDM chord via the cancellation process, which does not use slicing.

- EU-OFDM [16] (not plotted in Fig. 8) reports SNRs 1.5, 2, & 3-dB below their optimally biased DCO-OFDM for 64, 256 and 1024-QAM at 87.5% of the DCO's SE. In comparison, EACO-OFDM offers 5.5, 6.9 and 7.1-dB less cost than our optimum DCO-OFDM results for the same constellations and SE, as indicated on the right-hand-side of the graph. This is due to slicing errors at the lowest depth causing strong error vectors on the deeper depths, so that one bit-error may become three.

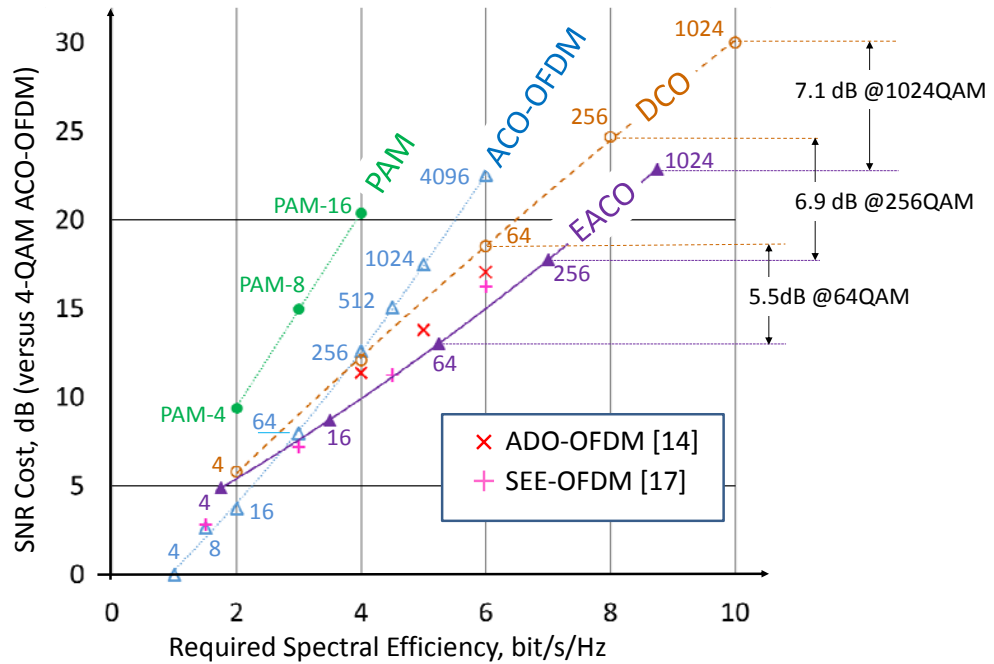

Fig. 8. Cost, in terms of electrical SNR, of using a higher spectral efficiency to support a fixed data rate at a fixed optical power, for ACO/EACO/DCO/ADO/SEE and EU-OFDM.

*3.6 Confirmation of the role of slicing in the cancellation process*

It is clear from Figs. 6 and 7 that the performances of the ACO-OFDM chords converge at higher signal to noise ratios. To confirm that this desirable effect is due to slicing, a simulation was run where the slicers were removed, and the cancellation mechanism became linear as in some previous schemes, such as by Dissanayaki and Armstrong [14]. Figure 9 plots the EVM for the three chords, with (solid lines) and without (dashed lines) slicing. Chord C0 (red) has identical performance, with and without slicing, as it does not use cancellation. Chord C1 benefits from slicing, even for low SNRs, but most strongly when the BER has fallen below approximately $10^{-3}$. The benefit of slicing for C2 is even stronger, simply because C2 required both C0 and C1 to be cancelled correctly. Note that, without slicing, the penalty in C1 and C2 relative to C0 appears constant across all plotted SNRs. This result confirms that slicing is a vital part of EACO-OFDM, especially if high constellation sizes are to be supported.

In EU-OFDM the slicing penalty is described by a scaling factor of the SNR (Equation 14, [16]). This translates to a constant offset if plotted on dB axes, as seen at our Fig. 8 for low SNRs. However, our system offers almost zero penalty at higher SNRs, implying that the scaling factor reduces to unity at high SNRs for EACO-OFDM.

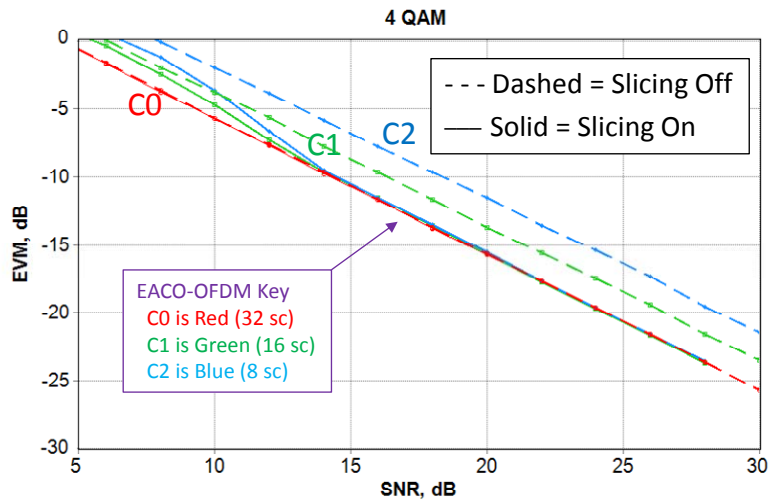

Fig. 9. Comparison of the performance of EACO-OFDM with and without the slicing operation in the cancellation process. The simulation parameters are as in Fig. 6. Without slicing, the higher chords (C1, green; C2, blue) suffer 2- and 4-dB penalties with respect to C0 (red) at all SNRs. With slicing, the penalties reduce to almost nothing above 15-dB SNR. Even at low SNRs, the slicing reduces these penalties.

## 4. Discussion

*4.1 Spectral efficiency considerations*

The results in Fig. 6 and 7 showed that ACO-OFDM outperforms DCO-OFDM in terms of the SNR required (essentially, for the same receiver noise) for a given BER at a given optical power. However, the spectral efficiency has been reduced to 87.5% of DCO-OFDM, meaning that 15%-higher-bandwidth electrical and opto-electronic components would be required to support the same data rates. The spectral efficiency could be improved by adding further chords – adding a fourth chord would give 60 subcarriers in total, the spectral efficiency would rise to 93.75%. The cost is an increased number of Fourier transforms, both at the receiver and at the transmitter, and some penalty due to the optical power required for this extra chord. However, the smaller number of subcarriers supported by these higher chords may make it more efficient to use look-up table and filters at the transmitters and receivers, rather than Fourier transforms. Alternatively, smaller-sized Fourier transforms could be used to generate the sparse subcarriers, with a simple frequency shifter at their outputs to place them at the correct frequencies.

*4.2 Comparison with iterative decoding in long-haul optical OFDM*

Iterative decoding to cancel *signal × signal* interference (SSI) has been used for long-haul direct-detection optical OFDM (DDO-OFDM), e.g. [23]. SSI occurs because the photodiode is a square-law device, where the photocurrent is the low-frequency part of the modulus of the optical field. In intensity modulated optical OFDM such as ACO and DCO, *signal × signal* is actually the wanted signal; however, in DDO-OFDM SSI causes unwanted intermodulation products, because the wanted signal is *signal* × (*optical carrier*). Although there are similarities between clipping-at-zero and square-law distortion products (i.e. strong even harmonics), a fundamental difference between DDO-OFDM and ACO-OFDM is the source of distortion products. In ACO-OFDM, the distortion is generated at the transmitter due to deliberate clipping: in DDO-OFDM the interference is generated at the receiver, by the photodiode, and is usually considered to be undesirable. That said, the techniques used in EACO-OFDM may be useful in DDO-OFDM.

## 5. Conclusions

A novel method of enhancing the spectral efficiency of Asymmetrically Clipped Optical OFDM (called Enhanced ACO-OFDM) has been presented. This relies on separately generating clipped ACO-OFDM chords, before combining them at the transmitter. The higher-number chords are collocated with the even-harmonic clipping distortion products of the lower chords. To avoid the clipping distortion of Chord ($c$) affecting the subcarriers in Chord ($c+1$), the clipping distortion of Chord ($c$) is predicted and subtracted from the signal sent to the processor for Chord ($c+1$). This process is repeated for $c = 0$ upwards. In contrast to some previous schemes, slicing is used during the cancellation process, and so provides near-equal signal qualities across all chords when the receivers are designed to provide BERs below $10^{-3}$. In contrast to EU-OFDM, EACO-OFDM places each subcarrier on its own unique frequency, thus takes advantage of the FTs of higher chords to reject the subcarriers of lower chords. This, along with the spreading of un-cancelled clipping noise amongst several subcarrier frequencies, reduces the effect of slicing errors of lower chords on the BER of higher chords.

Using ACO-OFDM signaling has advantages over Flip/U-OFDM-based schemes: (1) the signals are periodic within the common OFDM symbol's duration, so that no cyclic prefixes are required within the symbol, which reduces overheads; (2) a signal on a subcarrier frequency in the channel will uniquely appear at one output of the receiver's FT, enabling simple training and 1-tap equalization.

The 1024-QAM signal quality for a given number of subcarriers, optical power, and receiver noise, is at least 7-dB higher for EACO-OFDM than for DCO-OFDM, at 87.5% of DCO-OFDM's spectral efficiency, *at the same bit rate*. This figure assumes that the bias in the DCO-OFDM system is adaptively tuned to give optimum performance at a given signal-to-noise ratio at the receiver; for every 1-dB of over-bias, the advantage of ACO-OFDM increases by a further 1 dB. When compared with using higher-order ACO-OFDM to high spectral efficiencies, EACO-OFDM shows its greatest advantage at very high SNRs, which could support high-order QAM modulation. This advantage is a result of slicing during the cancellation process. High SNRs could be found in visible lightwave (VLC) systems, and so EACO-OFDM would benefit such systems considerably, by enabling high data rates over the bandwidth-limited channels typically found in VLC systems.

The proposed method has not made any use of sophisticated techniques for noise rejection in clipped-signal system, thus it is likely that the performance can be further improved by using the information content of the distortion products, rather than simply cancelling them. This is, of course, more difficult when the even frequencies also have information-bearing subcarriers upon them; however, strategies involving multiple iterations of cancellation may be worthwhile.

## Acknowledgments

I thank VPIphotonics (www.vpiphotonics.com) for the use of their simulator, VPItransmissionMakerWDM V9.1. This work is supported under the Australian Research Council's Laureate Fellowship scheme (FL130100041).